\documentclass[twocolumn,showpacs,preprintnumbers,amsmath,amssymb]{revtex4}
\usepackage{graphicx}
\usepackage{dcolumn}
\usepackage{bm}
\usepackage[colorlinks,linkcolor=blue]{hyperref}

\begin{document}

\preprint{APS/123-QED}

\title{Spin-isospin selectivity in three-nucleon forces}

\author{\firstname{H.}~\surname{Mardanpour$^{1}$}}
\author{\firstname{H.R.}~\surname{Amir-Ahmadi$^{1}$}}
\author{\firstname{R.}~\surname{Benard$^{1}$}}
\author{\firstname{A.}~\surname{Biegun$^{2,1}$}} 
\author{\firstname{M.}~\surname{Eslami-Kalantari$^{1,8}$}}
\author{\firstname{L.}~\surname{Joulaeizadeh$^{1}$}}
\author{\firstname{N.}~\surname{Kalantar-Nayestanaki$^{1}$}}
\author{\firstname{M.}~\surname{Ki\v{s}$^{1}$}}
\author{\firstname{St.}~\surname{Kistryn$^{3}$}}
\author{\firstname{A.}~\surname{Kozela$^{4}$}}
\author{\firstname{H.}~\surname{Kuboki$^{5}$}}
\author{\firstname{Y.}~\surname{Maeda$^{5}$}}
\author{\firstname{M.}~\surname{Mahjour-Shafiei$^{1,6}$}}
\author{\firstname{J.G.}~\surname{Messchendorp$^{1}$}}\email{messchendorp@kvi.nl}
\author{\firstname{K.}~\surname{Miki$^{5}$}}
\author{\firstname{S.}~\surname{Noji$^{5}$}}
\author{\firstname{A.}~\surname{Ramazani-Moghaddam-Arani$^{1,9}$}}
\author{\firstname{H.}~\surname{Sakai$^{5}$}}
\author{\firstname{M.}~\surname{Sasano$^{5}$}}
\author{\firstname{K.}~\surname{Sekiguchi$^{7}$}}
\author{\firstname{E.}~\surname{Stephan$^{2}$}}
\author{\firstname{R.}~\surname{Sworst$^{3}$}}
\author{\firstname{Y.}~\surname{Takahashi$^{5}$}} 
\author{\firstname{K.}~\surname{Yako$^{5}$.}}

\affiliation{
\vspace{2mm}
\begin{minipage}[t]{150mm}
$^{1}$KVI,~University~of~Groningen,~Groningen,~The~Netherlands,\\
$^{2}$Institute~of~Physics,~University~of~Silesia,~Katowice,~Poland,\\
$^{3}$Institute~of~Physics,~Jagellonian~University,~Krakow,~Poland,\\
$^{4}$Institute~of~Nuclear~Physics~PAN,~Krakow,~Poland,\\
$^{5}$Department~of~Physics,~University~of~Tokyo,~Tokyo,~Japan,\\
$^{6}$Department~of~Physics,~University~of~Tehran,~Tehran,~Iran,\\
$^{7}$RIKEN,~Tokyo,~Japan,\\
$^{8}$Department~of~Physics,~Faculty~of~Science,~Yazd~University,~Yazd,~Iran,\\
$^{9}$Department~of~Physics,~Faculty~of~Science,~University~of~Kashan,~Kashan,~Iran.
\end{minipage}
}
\date{\today}

\begin{abstract}
Precision data are presented for the break-up reaction, $^2{\rm
H}(\vec p,pp)n$, within the framework of nuclear-force studies. The
experiment was carried out at KVI using a polarized-proton beam of 
190~MeV impinging on a liquid-deuterium target and by exploiting the
detector, BINA. Some of the vector-analyzing powers are presented and
compared with state-of-the-art Faddeev calculations including
three-nucleon forces effect. Significant discrepancies between the
data and theoretical predictions were observed for kinematical
configurations which correspond to the $^2{\rm H}(\vec p,^2$He$)n$
channel. These results are compared to the $^2{\rm H}(\vec p,d)p$
reaction to test the isospin sensitivity of the present three-nucleon
force models. The current modeling of two and three-nucleon forces
is not sufficient to describe consistently polarization data for both 
isospin states.
\end{abstract}

\pacs{
{21.30.-x}{     Nuclear forces }, 
{21.45.+v}{     Few-body systems}, 
{24.70.+s}{     Polarization phenomena in reactions},
{25.45.De}{     Elastic and inelastic scattering}
}
\keywords{Suggested keywords}
\maketitle

Understanding the exact nature of the nuclear force is one of the
long-standing questions in nuclear physics. In 1935, Yukawa
successfully described the pair-wise nucleon-nucleon (NN) interaction
as an exchange of a boson~\cite{Yukawa}. Current NN models are mainly
based on Yukawa's idea and provide an excellent description of the
high-quality database of proton-proton and neutron-proton
scattering~\cite{stoks94} and of the properties of the
deuteron. However, for the simplest three-nucleon system, triton,
three-body calculations employing NN forces clearly underestimate the
experimental binding energies~\cite{wiringa95}, demonstrating
that NN forces are not sufficient to describe the three-nucleon system
accurately. Some of the discrepancies between experimental data and
calculations solely based on the NN interaction can be resolved by
introducing an additional three-nucleon force (3NF). Most of the
current models for the 3NF are based on a refined version of
Fujita-Miyazawa's 3NF model~{\cite{fuji}}, in which a 2$\pi$-exchange
mechanism is incorporated by an intermediate $\Delta$ excitation of
one of the nucleons~{\cite{deltuva03II,coon01}}.

The structure of the 3NF can be studied via a measurement of observables
in three-nucleon scattering processes. More detailed information on the 
spin dependence of the 3NF can be obtained by measuring polarization 
observables such as the analyzing powers. For this, a series of extensive 
studies of 3NF effects in elastic-scattering reactions have been performed 
at KVI and other laboratories. Precision measurements of the vector analyzing 
power of the proton in elastic proton-deuteron scattering have been performed
at various beam energies ranging from 90 to 250~MeV~{\cite{ermisch05,bieber,ermisch01,hatana,Ahmad08}}. 
Also, vector and tensor analyzing powers in elastic deuteron-proton scattering have
been obtained at various beam energies ranging from 75 to 
270~MeV~{\cite{witala,garcon,mardanEPJA07,ela07,kimiko,cadman}.
In these measurements, systematic discrepancies between data and 
theoretical predictions which rigorously solve the Faddeev equations and 
using only NN potentials were observed. A large part of the discrepancies were
removed by adding a 3NF to the NN potentials. Nevertheless, there are
still unresolved problems specially at higher energies, above
150~MeV/nucleon, which led to a request of more detailed
investigations. So far, none of the existing precision calculations
has produced a consistent explanation for all the experimental observables
in the intermediate energy range.

\begin{figure*}
\centering
\includegraphics[angle=0,width=.8\textwidth]{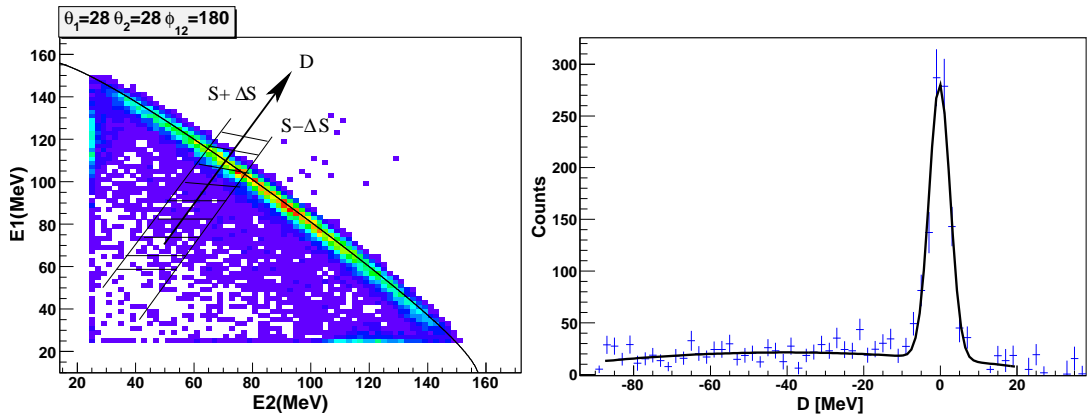}
\vspace*{-0.4cm}
\caption{The left panel shows the energy correlation between the 
 two protons for the kinematical configuration
 $(\theta_{1},\theta_{2},\phi_{12})$ =
 $(28^{\circ},28^{\circ},180^{\circ})$, together with the kinematical
 $S$-curve. In the right panel, a projection of events from a sample
 gate indicated in the left panel, $S=150\pm 4$ MeV, onto an axis D
 perpendicular to the $S$-curve is shown as crosses.  The solid line
 depicts a fit to that spectrum, composed of a Gaussian and a
 polynomial background model.}
\label{E1E2}
\end{figure*}

Complementary to the elastic scattering experiments, three-nucleon
studies have been performed exploiting the proton-deuteron break-up
reaction. The phase space of the break-up channel is much richer than that of
the elastic scattering. The final state of the break-up reaction is
described by 5 kinematical variables, as compared to just one for the
elastic scattering case. Therefore, studies of the break-up reaction
offer a way of much more detailed investigations of the nuclear
forces, in particular of the role of 3NF effects. Predictions show that 
large 3NF effects can be expected at specific kinematical regions in the
break-up reaction. Results of the cross sections and tensor analyzing
powers have already been published for a deuteron-beam energy of 130~MeV 
on a liquid-hydrogen target~{\cite{kistryn06,ola06,ela07}}. These
experiments were the first ones of its type which demonstrated the 
feasibility of a high-precision measurement of the break-up observables and they
confirmed that sizable influences of 3NF and Coulomb effects are
visible in the break-up cross sections at this energy. In the last
years, more data at several beam energies and other observables have
been collected to provide an extensive database at intermediate
energies. Here, we report on results obtained at relatively large
energies.

The behavior of the 3NF effects at higher energies has been
investigated via the $\vec{p}+d$ break-up reaction. The experiment was
performed at KVI using a polarized proton beam with an energy of
190~MeV impinging on a liquid-deuterium target. The reaction channel
has been identified using a 4$\pi$, highly symmetric detector system
\textbf{B}ig \textbf{I}nstrument for
\textbf{N}uclear-polarization \textbf{A}nalysis abbreviated as
\textbf{BINA}~\cite{mardanNPA07,mardanThesis,Ahmad08}. The relatively high 
energy used in this experiment offered a unique chance to study 3NF
effects, since their magnitude are predicted to increase with energy.  
In this paper, we present a set of selected analyzing power results,
preceded by a brief description of the methods used in the data analysis. We focus
specifically on results of the analyzing power at symmetric
configurations including those with very small azimuthal opening angles. Results are
compared with predictions of the modern Faddeev calculations.

The $\vec p+d$ reaction can lead to elastic scattering,
$\vec{p}+d\rightarrow p+d$, and the break-up of a deuteron
$\vec{p}+d\rightarrow p+p+n$. In the present experiment, both
processes have been observed and analyzed. The differential cross
section and analyzing power of the elastic channel have been measured
at KVI before~\cite{ermisch05,ermisch01}. The values of the elastic
scattering cross sections evaluated from the current data agree very
well with the previous measurements. The polarization of the proton beam
was extracted by exploiting the cross section asymmetry of the elastic
scattering reaction and by using the analyzing power from the previous
measurements at KVI. These values were used in the determination of the
analyzing power for the break-up channel. In addition, the elastic
channel is used for the energy calibration and for monitoring of the
whole setup with respect to time drifts.

Conventionally, in the $\vec{p}+d$ break-up reaction, the kinematics
are determined by using the scattering angles of the two final-state
protons, $(\theta_{1},\theta_{2},\phi_{12}=\phi_{1}-\phi_{2})$ where
$\theta_{1},\theta_{2}$ are the polar scattering angles of the first
and second proton, respectively, and $\phi_{12}$ is the azimuthal
angle between the two protons. The left panel in Fig.~\ref{E1E2} shows
the correlation between the energies of the two protons for a sample
geometry, namely $(\theta_{1},\theta_{2},\phi_{12})$ =
$(28^{\circ},28^{\circ},180^{\circ})$.  The expected correlation
according to the relativistic kinematics for the break-up reaction,
referred to as the $S$-curve, is shown as the solid line. The
kinematical variable, $S$, is defined as the arc-length along this
curve, starting from the minimum value of $E_{1}$. It is customary to
present the cross sections and analyzing powers as a function of the
variable $S$. The right panel in Fig.~\ref{E1E2} depicts a projection of
the spectrum onto an axis D perpendicular to the $S$-curve and for a 
window of $\pm \Delta S$. The low-energy tail corresponds predominantly 
to events in which one of the protons of the break-up reaction undergoes 
a hadronic interaction inside the plastic scintillator of BINA, thereby depositing only a
fraction of its energy. The background from other sources, such as
time-uncorrelated pile-up events, were found to be negligible.

\begin{figure}[h]
\centering
\includegraphics[angle=0,width=.48\textwidth]
{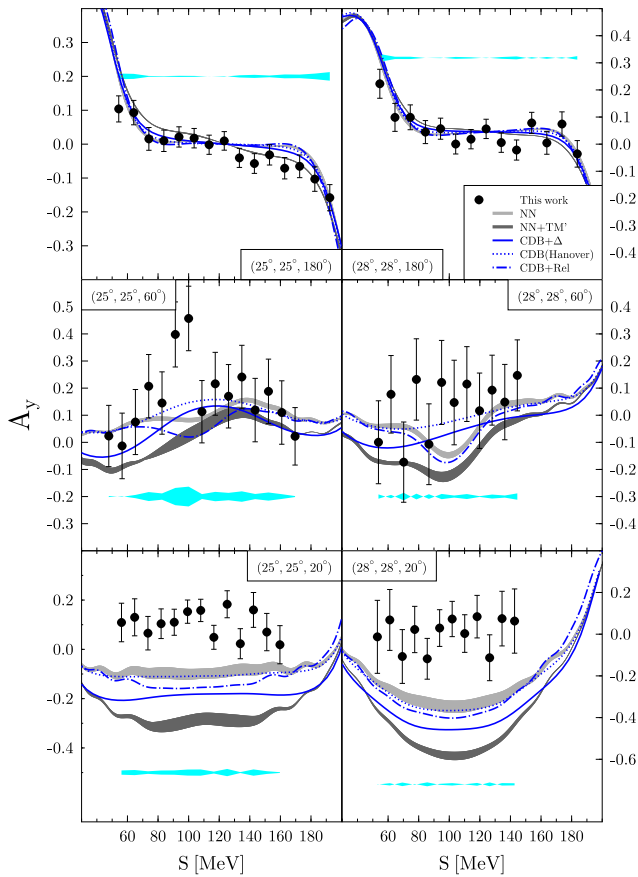}
\vspace*{-0.8cm}
\caption{A comparison between the results of the analyzing power
 measurements for a few selected break-up configurations with various
 theoretical predictions. The light gray bands are composed of various
 modern two-nucleon (NN) force calculations, namely CD-Bonn, NijmI,
 NijmII, and AV18. The dark gray bands correspond to results of the
 calculations with the same NN forces including the TM' (3N)
 potential.  The lines represent the predictions of calculations by
 the Hannover-Lisbon group based on the CD-Bonn potential (dotted) and
 CD-Bonn potential extended with a virtual $\Delta$ excitation (solid blue). 
 The blue dash-dotted lines are derived from calculations by
 the Bochum-Cracow collaboration based on the CD-Bonn potential
 including relativistic effects~\cite{skib06}. The errors are statistical and the
 cyan band in each panel represents the systematic uncertainties
 (2$\sigma$).}
\label{crossall}
\end{figure}

The interaction of a polarized beam with an unpolarized target
produces an azimuthal asymmetry in the scattering cross section. BINA
has a complete azimuthal coverage and can, therefore, unambiguously
determine the magnitude of the asymmetry,
$\frac{\sigma^{\downarrow}-\sigma^{\uparrow}}{\sigma^{\uparrow}p_{Z}^{\downarrow}-\sigma^{\downarrow}p_{Z}^{\uparrow}}$
with $p_{Z}$ the polarization of the incident beam and
$\sigma^{\downarrow}$, $\sigma^{\uparrow}$ the 5-fold differential
cross section in the case of beam polarizations pointing ``down'' and ``up'', 
respectively. This asymmetry corresponds to
$A_{y}\cdot\cos(\phi)$ with $\phi$ the azimuthal angle of the reaction
plane, understood as the plane spanned by the momentum vectors of the
beam and of the ``first'' emitted proton, and $A_y$ the vector analyzing
power. Note that, in first order, the polarization observable, $A_y$,
does not suffer from uncertainties in detection efficiencies and
acceptances, since these cancel out in the calculation of this observable. The dominant
part of the systematic uncertainty stems from the polarizations
$p_{Z}^{\uparrow}$ and $p_{Z}^{\downarrow}$. The beam polarization was
determined independently via asymmetry measurements of the elastic
proton-proton scattering process using the in-beam polarimeter,
IBP~\cite{Bieber01} and asymmetry measurements of the elastic
proton-deuteron scattering using BINA. During the experiment, the
proton polarization was typically 60\% for $p_{Z}^{\uparrow}$ and
$p_{Z}^{\downarrow}$.

Figure~\ref{crossall} presents results of the analyzing powers for two
symmetric kinematical configurations $(\theta_{1},\theta_{2})$=
($25^{\circ},25^{\circ}$) and ($28^{\circ},28^{\circ}$) for three
different values of $\phi_{12}$. The data are compared with
calculations based on different models for the interaction dynamics as
described in detail in the caption of the figure. For these configurations and
observable, the effects of relativity and the Coulomb force are
predicted to be small with respect to the effect of three-nucleon
forces.  At $\phi_{12}$=$180^\circ$, the value of $A_y$ is predicted
to be completely determined by two-nucleon force effects with only a
very small effect of 3NFs, which is supported by the experimental
data. Note, however, that the effect of 3NFs increases with decreasing
of the relative azimuthal angle $\phi_{12}$, corresponding to a
decrease in the relative energy between the two final-state protons.

\begin{figure}[h]
\centering
\includegraphics[angle=0,width=.48\textwidth]{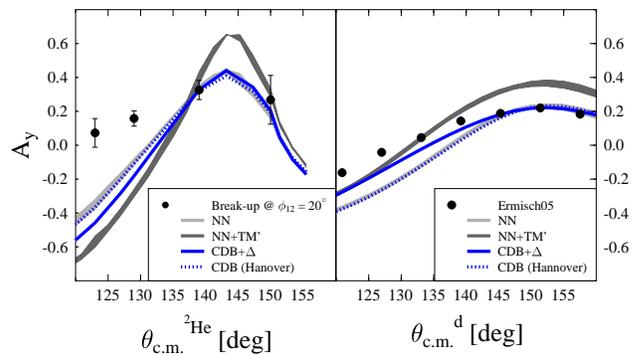}
\vspace*{-0.8cm}
\caption{The analyzing power as a function of the center of mass
 angle for two reactions $^2{\rm H}(\vec p,^2$He$)n$ (left panel) and
 $^2{\rm H}(\vec p,d)p$ (right panel). For a description of the lines
 and bands, see Fig.~\ref{crossall}. The data of the $^2{\rm
 H}(\vec p,d)p$ reaction are taken from
 Refs.~\cite{ermisch05,ermisch01}.}
\label{he2ay}
\end{figure}

A surprising discrepancy between the measured analyzing powers and
theoretical predictions can be observed at small relative azimuthal
opening angles $\phi_{12}$$=$$20^{\circ}$. This configuration corresponds
to a relative energy between the two protons of less than 10~MeV.
Note that this deficiency even increases when including three-nucleon
force effects such as the TM' potential or the implicit inclusion of
the $\Delta$ isobar by the Hannover-Lisbon theory group.  The relative
energy between the two protons varies as a function of $S$ and for
symmetric configurations, $\theta_{1}=\theta_{2}$, it reaches a very low
value at the center of $S$ of less than 1~MeV. In these cases, the two
protons move very close to each other in a relative angular momentum $S$ state 
with an isospin of one, which is similar to the configuration of a $^2$He. The
analyzing power for the corresponding reaction, $^2{\rm H}(\vec
p,^2$He$)n$, can be compared to the analyzing power of the elastic
$^2{\rm H}(\vec p,d)p$ reaction. In the elastic channel, the total
isospin of the initial and final state is exclusively 1/2, whereas in
the former case, the final state might couple to an isospin 3/2 as a
consequence of the isospin violating Coulomb force. For a comparison,
we extracted the analyzing power, $A_y$, for the $^2$He state at a kinematics
corresponding to a value in the middle of the $S$-curve where the relative energy is at
its minimum. Figure~\ref{he2ay} depicts the resulting analyzing power
as a function of the center-of-mass angle for the two reactions
$^2{\rm H}(\vec p,^2$He$)n$ (left panel) and $^2{\rm H}(\vec p,d)p$
(right panel). For the left-hand side panel, the center-of-mass angles 
are obtained by assuming that the $pp$ pair with a small relative energy
corresponds to one body, namely a $^2$He state scattering at an angle 
$\theta_1$=$\theta_2$. The theory curves depicted in the right panel were
obtained from calculations for the symmetric configuration of the
break-up reaction, ($\theta_{1}=\theta_{2}$, $\phi_{12} = 20^{\circ}$), 
and taken at the center of $S$. Note that at center-of-mass angles 
of less than 135$^\circ$, there
is a large discrepancy between the state-of-art calculations and the
experimental data for the $^2{\rm H}(\vec p,^2$He$)n$ reaction,
whereas the same calculations deviate significantly less with the
analyzing power results in the $^2{\rm H}(\vec p,d)p$ channel at the
same incident energy. The current modeling of two and three-nucleon
forces is not sufficient to describe consistently polarization data
for the two isospin states, which hints towards a deficiency in the 
spin-isospin structure of the forces.

This paper presents a study of the vector analyzing power, $A_y$, in
the proton-deuteron break-up reaction with an incident proton-beam
energy of 190~MeV.  The data were obtained exploiting a detection
system, BINA, which covers nearly the full kinematical phase space of
the break-up reaction. In particular, it features a complete azimuthal
coverage which provides a well-controlled measure of spin observables. 
The analyzing power, $A_y$, has proven to be a
unique probe to study 3NF effects, especially since the effect of the
Coulomb force and relativity are expected to be small. For kinematical
configurations at which the relative azimuthal opening angle between
the two final-state protons is small, 3NF effects are predicted to be
large. These regions in phase space can, therefore, be studied
rigorously by a comparison with experimental data. Here, we
concentrate on the spin-isospin structure of the three-nucleon force,
which can be tested by a comparison between the break-up channel
mimicking the, $^2{\rm H}(\vec p,^2$He$)n$ reaction, and the elastic
scattering channel, $^2{\rm H}(\vec p,d)p$. For this, we analyzed the
break-up reaction at configurations at which the relative energy
between the two final-state protons is at its minimum within the
experimental acceptance. Strikingly, Faddeev calculations, which are
based on modern two and three-nucleon potentials, fail to describe the
analyzing powers in the $^2{\rm H}(\vec p,^2$He$)n$ channel, whereas
the same calculations compare well to polarization data in the
analogous elastic channel. Such an inconsistency points to a
deficiency in the spin-isospin structure of the description of the
many-nucleon forces in the present-day state-of-the-art calculations.

The authors acknowledge the work by the cyclotron and ion-source
groups at KVI for delivering a high-quality beam used in these
measurements. We thank Arnoldas Deltuva, Peter Sauer, and Henryk
Wita\l{a} for valuable discussions, the theoretical basis and
corresponding calculations. This work was performed as part of the
research program of the ``Stichting voor Fundamenteel Onderzoek der
Materie'' (FOM). Furthermore, the present work has been performed 
with financial support from the University of Groningen (RuG) 
and the Helmholtzzentrum f\"ur Schwerionenforschung GmbH (GSI), 
Darmstadt.

\bibliography{PRL_pdbreakup_190}

\begin{thebibliography}{23}
\expandafter\ifx\csname natexlab\endcsname\relax\def\natexlab#1{#1}\fi
\expandafter\ifx\csname bibnamefont\endcsname\relax
  \def\bibnamefont#1{#1}\fi
\expandafter\ifx\csname bibfnamefont\endcsname\relax
  \def\bibfnamefont#1{#1}\fi
\expandafter\ifx\csname citenamefont\endcsname\relax
  \def\citenamefont#1{#1}\fi
\expandafter\ifx\csname url\endcsname\relax
  \def\url#1{\texttt{#1}}\fi
\expandafter\ifx\csname urlprefix\endcsname\relax\def\urlprefix{URL }\fi
\providecommand{\bibinfo}[2]{#2}
\providecommand{\eprint}[2][]{\url{#2}}

\bibitem[{\citenamefont{Yukawa}(1935)}]{Yukawa}
\bibinfo{author}{\bibfnamefont{H.}~\bibnamefont{Yukawa}},
  \bibinfo{journal}{Proc. Phys. Math. Soc. Jap.} \textbf{\bibinfo{volume}{17}},
  \bibinfo{pages}{48} (\bibinfo{year}{1935}).

\bibitem[{\citenamefont{Stoks et~al.}(1994)\citenamefont{Stoks, Klomp,
  Terheggen, and de~Swart}}]{stoks94}
\bibinfo{author}{\bibfnamefont{V.~G.~J.} \bibnamefont{Stoks}},
  \bibinfo{author}{\bibfnamefont{R.~A.~M.} \bibnamefont{Klomp}},
  \bibinfo{author}{\bibfnamefont{C.~P.~F.} \bibnamefont{Terheggen}},
  \bibnamefont{and} \bibinfo{author}{\bibfnamefont{J.~J.}
  \bibnamefont{de~Swart}}, \bibinfo{journal}{Phys. Rev. C}
  \textbf{\bibinfo{volume}{49}}, \bibinfo{pages}{2950} (\bibinfo{year}{1994}).

\bibitem[{\citenamefont{Wiringa et~al.}(1995)\citenamefont{Wiringa, Stoks, and
  Schiavilla}}]{wiringa95}
\bibinfo{author}{\bibfnamefont{R.~B.} \bibnamefont{Wiringa}},
  \bibinfo{author}{\bibfnamefont{V.~G.~J.} \bibnamefont{Stoks}},
  \bibnamefont{and}
  \bibinfo{author}{\bibfnamefont{R.}~\bibnamefont{Schiavilla}},
  \bibinfo{journal}{Phys. Rev. C} \textbf{\bibinfo{volume}{51}},
  \bibinfo{pages}{38} (\bibinfo{year}{1995}).

\bibitem[{\citenamefont{Fujita and Miyazawa}(1957)}]{fuji}
\bibinfo{author}{\bibfnamefont{J.}~\bibnamefont{Fujita}} \bibnamefont{and}
  \bibinfo{author}{\bibfnamefont{H.}~\bibnamefont{Miyazawa}},
  \bibinfo{journal}{Prog. Theor. Phys.} \textbf{\bibinfo{volume}{17}},
  \bibinfo{pages}{360} (\bibinfo{year}{1957}).

\bibitem[{\citenamefont{Deltuva et~al.}(2003)\citenamefont{Deltuva, Machleidt,
  and Sauer}}]{deltuva03II}
\bibinfo{author}{\bibfnamefont{A.}~\bibnamefont{Deltuva}},
  \bibinfo{author}{\bibfnamefont{R.}~\bibnamefont{Machleidt}},
  \bibnamefont{and} \bibinfo{author}{\bibfnamefont{P.~U.} \bibnamefont{Sauer}},
  \bibinfo{journal}{Phys. Rev. C} \textbf{\bibinfo{volume}{68}},
  \bibinfo{pages}{024005} (\bibinfo{year}{2003}).

\bibitem[{\citenamefont{Coon and Han}(2001)}]{coon01}
\bibinfo{author}{\bibfnamefont{S.~A.} \bibnamefont{Coon}} \bibnamefont{and}
  \bibinfo{author}{\bibfnamefont{H.~K.} \bibnamefont{Han}},
  \bibinfo{journal}{Few-Body Sys.} \textbf{\bibinfo{volume}{30}},
  \bibinfo{pages}{131} (\bibinfo{year}{2001}).

\bibitem[{\citenamefont{Ermisch et~al.}(2005)}]{ermisch05}
\bibinfo{author}{\bibfnamefont{K.}~\bibnamefont{Ermisch}} \bibnamefont{et~al.},
  \bibinfo{journal}{Phys. Rev. C} \textbf{\bibinfo{volume}{71}},
  \bibinfo{pages}{064004} (\bibinfo{year}{2005}).

\bibitem[{\citenamefont{Bieber et~al.}(2000)}]{bieber}
\bibinfo{author}{\bibfnamefont{R.}~\bibnamefont{Bieber}} \bibnamefont{et~al.},
  \bibinfo{journal}{Phys. Rev. Lett.} \textbf{\bibinfo{volume}{84}},
  \bibinfo{pages}{606} (\bibinfo{year}{2000}).

\bibitem[{\citenamefont{Ermisch et~al.}(2001)}]{ermisch01}
\bibinfo{author}{\bibfnamefont{K.}~\bibnamefont{Ermisch}} \bibnamefont{et~al.},
  \bibinfo{journal}{Phys. Rev. Lett.} \textbf{\bibinfo{volume}{86}},
  \bibinfo{pages}{5862} (\bibinfo{year}{2001}).

\bibitem[{\citenamefont{Hatanaka et~al.}(2002)}]{hatana}
\bibinfo{author}{\bibfnamefont{K.}~\bibnamefont{Hatanaka}}
  \bibnamefont{et~al.}, \bibinfo{journal}{Phys. Rev. C}
  \textbf{\bibinfo{volume}{66}}, \bibinfo{pages}{044002}
  (\bibinfo{year}{2002}).

\bibitem[{\citenamefont{Ramazani-Moghaddam-Arani et~al.}(2008)}]{Ahmad08}
\bibinfo{author}{\bibfnamefont{A.}~\bibnamefont{Ramazani-Moghaddam-Arani}}
  \bibnamefont{et~al.}, \bibinfo{journal}{Phys. Rev. C}
  \textbf{\bibinfo{volume}{78}}, \bibinfo{pages}{014006(R)}
  (\bibinfo{year}{2008}).

\bibitem[{\citenamefont{Wita\l{a} et~al.}(1993)\citenamefont{Wita\l{a},
  Cornelius, and Gl{\"o}ckle}}]{witala}
\bibinfo{author}{\bibfnamefont{H.}~\bibnamefont{Wita\l{a}}},
  \bibinfo{author}{\bibfnamefont{T.}~\bibnamefont{Cornelius}},
  \bibnamefont{and}
  \bibinfo{author}{\bibfnamefont{W.}~\bibnamefont{Gl{\"o}ckle}},
  \bibinfo{journal}{Few-Body Sys.} \textbf{\bibinfo{volume}{15}},
  \bibinfo{pages}{67} (\bibinfo{year}{1993}).

\bibitem[{\citenamefont{Gar\c{c}on et~al.}(1986)}]{garcon}
\bibinfo{author}{\bibfnamefont{M.}~\bibnamefont{Gar\c{c}on}}
  \bibnamefont{et~al.}, \bibinfo{journal}{Nucl. Phys. A}
  \textbf{\bibinfo{volume}{458}}, \bibinfo{pages}{287} (\bibinfo{year}{1986}).

\bibitem[{\citenamefont{Mardanpour et~al.}(2007{\natexlab{a}})}]{mardanEPJA07}
\bibinfo{author}{\bibfnamefont{H.}~\bibnamefont{Mardanpour}}
  \bibnamefont{et~al.}, \bibinfo{journal}{Eur. Phys. J. A}
  \textbf{\bibinfo{volume}{31}}, \bibinfo{pages}{383}
  (\bibinfo{year}{2007}{\natexlab{a}}).

\bibitem[{\citenamefont{Stephan et~al.}(2007)}]{ela07}
\bibinfo{author}{\bibfnamefont{E.}~\bibnamefont{Stephan}} \bibnamefont{et~al.},
  \bibinfo{journal}{Phys.~Rev.~C} \textbf{\bibinfo{volume}{76}},
  \bibinfo{pages}{057001} (\bibinfo{year}{2007}).

\bibitem[{\citenamefont{Sekiguchi et~al.}(2002)}]{kimiko}
\bibinfo{author}{\bibfnamefont{K.}~\bibnamefont{Sekiguchi}}
  \bibnamefont{et~al.}, \bibinfo{journal}{Phys. Rev. C}
  \textbf{\bibinfo{volume}{65}}, \bibinfo{pages}{034003}
  (\bibinfo{year}{2002}).

\bibitem[{\citenamefont{Cadman et~al.}(2001)}]{cadman}
\bibinfo{author}{\bibfnamefont{R.~V.} \bibnamefont{Cadman}}
  \bibnamefont{et~al.}, \bibinfo{journal}{Phys. Rev. Lett.}
  \textbf{\bibinfo{volume}{86}}, \bibinfo{pages}{967} (\bibinfo{year}{2001}).

\bibitem[{\citenamefont{Kistryn et~al.}(2006)}]{kistryn06}
\bibinfo{author}{\bibfnamefont{S.}~\bibnamefont{Kistryn}} \bibnamefont{et~al.},
  \bibinfo{journal}{Phys. Lett. B} \textbf{\bibinfo{volume}{641}},
  \bibinfo{pages}{23} (\bibinfo{year}{2006}).

\bibitem[{\citenamefont{Biegun et~al.}(2006)}]{ola06}
\bibinfo{author}{\bibfnamefont{A.}~\bibnamefont{Biegun}} \bibnamefont{et~al.},
  \bibinfo{journal}{Acta Phys. Pol.} \textbf{\bibinfo{volume}{B371}},
  \bibinfo{pages}{213} (\bibinfo{year}{2006}).

\bibitem[{\citenamefont{Mardanpour et~al.}(2007{\natexlab{b}})}]{mardanNPA07}
\bibinfo{author}{\bibfnamefont{H.}~\bibnamefont{Mardanpour}}
  \bibnamefont{et~al.}, \bibinfo{journal}{Nucl. Phys. A}
  \textbf{\bibinfo{volume}{790}}, \bibinfo{pages}{426c}
  (\bibinfo{year}{2007}{\natexlab{b}}).

\bibitem[{\citenamefont{Mardanpour}(2008)}]{mardanThesis}
\bibinfo{author}{\bibfnamefont{H.}~\bibnamefont{Mardanpour}}, Ph.D. thesis,
  \bibinfo{school}{University of Groningen} (\bibinfo{year}{2008}).

\bibitem[{\citenamefont{Skibi{\'n}ski et~al.}(2006)\citenamefont{Skibi{\'n}ski,
  Wita{\l}a, and Golak}}]{skib06}
\bibinfo{author}{\bibfnamefont{R.}~\bibnamefont{Skibi{\'n}ski}},
  \bibinfo{author}{\bibnamefont{Wita{\l}a}}, \bibnamefont{and}
  \bibinfo{author}{\bibfnamefont{J.}~\bibnamefont{Golak}},
  \bibinfo{journal}{Eur. Phys. J. A} \textbf{\bibinfo{volume}{30}},
  \bibinfo{pages}{369} (\bibinfo{year}{2006}).

\bibitem[{\citenamefont{Bieber et~al.}(2001)}]{Bieber01}
\bibinfo{author}{\bibfnamefont{R.}~\bibnamefont{Bieber}} \bibnamefont{et~al.},
  \bibinfo{journal}{Nucl. Instr. Meth. Phys. Res. A}
  \textbf{\bibinfo{volume}{457}}, \bibinfo{pages}{12} (\bibinfo{year}{2001}).

\end{thebibliography}
\end{document}